\documentclass[a4paper, amsfonts, amssymb, amsmath, reprint, showkeys, twoside,superscriptaddress, onecolumn, nofootinbib]{revtex4-2}
\usepackage{tikz}
\usetikzlibrary{quantikz2}
\usepackage{orcidlink}
\usepackage{graphicx}
\usepackage[left=0.6in,right=0.6in,top=1in,columnsep=15pt]{geometry} 
\usepackage{amsmath}
\usepackage{amsthm}
\usepackage{amssymb}
\usepackage{mathtools}
\usepackage{bm}
\usepackage{comment}
\usepackage{mathalpha}
\usepackage{xcolor}
\usepackage{amsfonts}
\usepackage{nicematrix}
\usepackage{hyperref}
\usepackage[capitalize,nameinlink]{cleveref}
\usepackage{thm-restate}
\usepackage{algorithm}
\usepackage{algpseudocode}
\usepackage{mhchem}
\usepackage{multirow}
\usepackage{makecell}
\usepackage{soul}
\usepackage{fontawesome}

\newif\ifmarkenabled

\markenabledfalse % or \markenabledfalse

\newcommand{\marktext}[1]{%
  \ifmarkenabled%
    \textcolor{red}{#1}%
  \else%
    \textcolor{black}{#1}%
  \fi%
}

\newtheorem{theorem}{Theorem}

\begin{document}
\title{Quantum Algorithm for Vibronic Dynamics: Case Study on Singlet Fission Solar Cell Design}

\author{Danial Motlagh \orcidlink{0009-0003-7655-4341}}
\email{danial.motlagh@xanadu.ai}
\affiliation{Xanadu, Toronto, ON, M5G 2C8, Canada}
\author{Robert A. Lang \orcidlink{0000-0002-4345-3566}}
\email{robert.lang@xanadu.ai}
\affiliation{Xanadu, Toronto, ON, M5G 2C8, Canada}
\author{Paarth Jain \orcidlink{0009-0005-9107-4789}}
\affiliation{Xanadu, Toronto, ON, M5G 2C8, Canada}
\affiliation{University of Toronto, Toronto, ON, M5S 3H6, Canada}
\author{\mbox{Jorge A. Campos-Gonzalez-Angulo \orcidlink{0000-0003-1156-3012}}}
\affiliation{University of Toronto, Toronto, ON, M5S 3H6, Canada}
\affiliation{Vector Institute for Artificial Intelligence, Toronto, ON, M5G 1M1, Canada}
\author{William Maxwell \orcidlink{0009-0005-8603-2955}}
\affiliation{Xanadu, Toronto, ON, M5G 2C8, Canada}
\author{Tao Zeng \orcidlink{0000-0002-1553-7850}}
\affiliation{Department of Chemistry, York University, Toronto, ON, M3J 1P3, Canada}
\author{Alan Aspuru-Guzik \orcidlink{0000-0002-8277-4434}}
\affiliation{University of Toronto, Toronto, ON, M5S 3H6, Canada}
\affiliation{Vector Institute for Artificial Intelligence, Toronto, ON, M5G 1M1, Canada}
\affiliation{Canadian Institute for Advanced Research (CIFAR), Toronto, ON, M5G 1M1, Canada}
\affiliation{Acceleration Consortium, Toronto, ON, M5S 3H6, Canada}
\author{Juan Miguel Arrazola \orcidlink{0000-0002-0619-9650}}
\affiliation{Xanadu, Toronto, ON, M5G 2C8, Canada}

% % Manually add the footnote
% \makeatletter
% \patchcmd{\@maketitle}
%   {\end{center}}
%   {%
%    \end{center}%
%    \renewcommand\thefootnote{\fnsymbol{footnote}}%
%    \footnotetext[1]{Corresponding author: \href{mailto:danial.motlagh@xanadu.ai}{\texttt{danial.motlagh@xanadu.ai}}}%
%    \footnotetext[2]{Corresponding author: \href{mailto:robert.lang@xanadu.ai}{\texttt{robert.lang@xanadu.ai}}}%
%   }
%   {}{}
% \makeatother

\begin{abstract}
Vibronic interactions between nuclear motion and electronic states are critical for the accurate modeling of photochemistry. However, accurate simulations of fully quantum non-adiabatic dynamics are often prohibitively expensive for classical methods beyond small systems. In this work, we present a quantum algorithm based on product formulas for simulating time evolution under a general vibronic Hamiltonian in real space, capable of handling an arbitrary number of electronic states and vibrational modes. We develop the first trotterization scheme for vibronic Hamiltonians beyond two electronic states and introduce an array of optimization techniques for the exponentiation of each fragment in the product formula, resulting in a remarkably low cost of implementation. To demonstrate practical relevance, we outline a proof-of-principle integration of our algorithm into a materials discovery pipeline for designing more efficient singlet fission-based organic solar cells. {We estimate that $100$ femtoseconds of propagation using a second-order Trotter product formula for a $6$-state, $21$-mode model of exciton transport at an anthracene dimer requires $154$ qubits and $2.76 \times 10^6$ Toffoli gates. While a $4$-state, $246$-mode model describing charge transfer at an anthracene-fullerene interface requires $1053$ qubits and $2.66 \times 10^7$ Toffoli gates.} 
\end{abstract}

\maketitle
\section{Introduction}

Much of the effort in applying quantum computers to problems in quantum chemistry has been concerned with electronic structure, particularly with the computation of ground state energies of the electronic Hamiltonian~\cite{cao2019quantum}. While of great importance, electronic structure on its own is not a complete description of molecular behavior, as it neglects the crucial role of vibronic effects caused by nuclear vibrational motions interacting with electronic states. Furthermore, it remains unclear whether electronic structure is where quantum advantage will be found~\cite{Lee2023}. Accurate vibronic simulations are essential for understanding photo-induced processes, such as non-radiative relaxation and energy transfer, and are instrumental in interpreting experimental spectra~\cite{mai2020molecular, Yarkony2012}. Studying such processes has many applications across a wide range of technologies. These include designing advanced materials for optoelectronic and photovoltaic technologies~\cite{long2017nonadiabatic, nelson2014nonadiabatic}, studying molecular photomagnets for ultradense memory storage~\cite{goodwin2017molecular, reta2021ab, staab2022analytic, Penfold_2023, mattioni2024vibronic}, enhancing the performance of thermally activated delayed fluorescence (TADF)-based organic light-emitting diodes (OLEDs)~\cite{Endo2011, Uoyama2012, Penfold2018, eng2021open}, discovering improved sunscreens~\cite{li2016photodynamics, wu2021nonadiabatic, zhao2021non, wang2022ultrafast}, and advancing photodynamic therapies for cancer treatment~\cite{via2001photochemotherapy, Zhu_Finlay2008, Cui_Fang2013, ponte2022computations}. Moreover, vibronic effects shed light on photo-excited biological processes, such as those in biological light-harvesting complexes~\cite{Neville_Schuurman2024, jaiswal2024sub} and DNA~\cite{barbatti2010relaxation, markwick2007ultrafast, green2021nonadiabatic}. However, despite their importance, full non-adiabatic dynamics are too resource-intensive for classical methods and are often omitted in materials discovery pipelines.\\

Many excited-state scenarios encountered in photochemistry, such as conical intersections and avoided crossings, require a description beyond the Born-Oppenheimer approximation~\cite{domcke2004conical, bersuker2013jahn, meyer2009multidimensional}. There has been much effort in developing numerical methods for performing non-adiabatic dynamics. Semi-classical methods, like trajectory surface hopping~\cite{tully1990molecular, chapman1992classical, thompson1998modern, subotnik2016understanding} and Ehrenfest dynamics~\cite{li2005ab, billing1976semiclassical, billing1983use, billing2003quantum, meyera1979classical} can be efficient but fail to capture important quantum phenomena such as coherence, tunneling, and wavepacket branching. For molecular applications, the most ubiquitously employed method for propagation according to the time-dependent Schr\"{o}dinger equation is the multi-configurational time-dependent Hartree (MCTDH) method~\cite{meyer1990multi, beck2000multiconfiguration, meyer2009multidimensional}, along with related techniques~\cite{wang2003multilayer, wang2015multilayer, burghardt2008multimode, richings2015quantum}. However, these methods are approximate, their use requires high levels of expertise, and have unfavorable scaling with increased accuracy, limiting them to small systems and short propagation times~\cite{marquetand2016challenges, mukherjee2022simulations}. Quantum computers offer an alternative to overcoming these limitations  \cite{ollitrault2021molecular, schleich2024chemically, miessen2023quantum}; however, little attention has been given to this topic in quantum computing. Analog quantum simulators have gathered particular interest, but they currently face scalability issues for large-scale calculations~\cite{macdonell2021analog, kang2024seeking}. Alternative digital-based algorithms have been explored~\cite{ollitrault2020nonadiabatic}, but these results are limited in scope and not applicable to general vibronic Hamiltonians with an arbitrary number of diabatic states or arbitrary forms of mode interactions.\\

\begin{figure}[t!]
    \centering
    \vspace{-0.2cm}
    \includegraphics[width=0.75\linewidth]{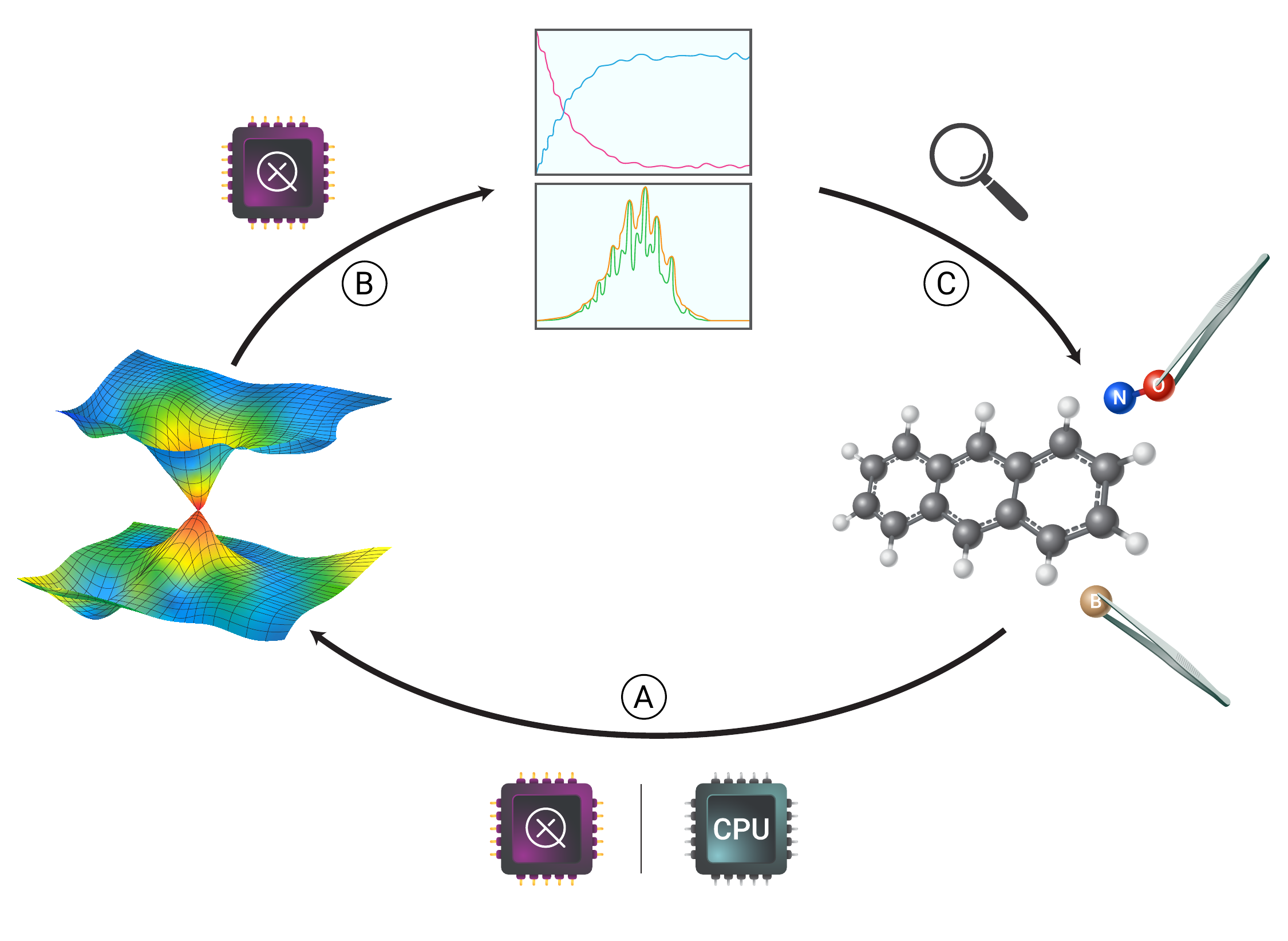}
    \vspace{-0.7cm}
    \caption{Proof-of-principle integration of our quantum algorithm into a materials discovery workflow. (A) Following the initialization or modification of a molecule, electronic structure calculations are performed, leveraging either quantum or classical resources to construct the vibronic model. For vibronic Hamiltonians employed in ultrafast photochemistry, this procedure typically includes geometry optimization, a normal-mode analysis following the calculation of the Hessian, and excited-state calculations at points along normal mode displacements. (B) The non-adiabatic dynamics for the vibronic Hamiltonian are performed on a quantum computer using the algorithm presented in Section~\ref{sec:algorithm}, producing observable quantities such as diabatic state populations and photoabsorption spectra. (C) The interpretation of the output observables is used to inform a change to molecular candidates to optimize properties such as transition rates and excited-state lifetimes.}
    \label{fig:workflow}
\end{figure}

In this work, we develop a quantum algorithm based on product formulas for time evolution under a general vibronic Hamiltonian with an arbitrary number of states and modes. One of the main technical novelties of our algorithm is a trotterization scheme for vibronic Hamiltonians beyond two electronic states. We also develop an array of optimization techniques deployed in the exponentiation of each fragment in the product formula to significantly reduce the cost of its implementation. After introducing the algorithm, we analyze its complexity and address initial state preparation relevant to photochemistry, showing that it incurs negligible cost compared to the simulation. Finally, we discuss the efficient extraction of observables specific to vibronic dynamics on a quantum computer. In particular, we show that extracting electronic state populations---the key observable in vibronic simulation---only requires measuring the few-qubit electronic register in the computational basis. This contrasts with typical energy-based observables requiring controlled implementations of the time evolution operator. Combined with cheap initial state preparation, this solves the input-output problem often faced in other applications of quantum computing.\\

Next, we discuss applications of our algorithm, with a particular focus on its relevance for organic solar cells based on the physical process of singlet fission (SF) \cite{smith2010singlet, smith2013recent, michl2019unconventional}, a form of multiple exciton generation that allows the production of two excitons from just one photon. This makes SF-based solar cells potentially capable of producing two pairs of free charge carriers from each photon, enhancing the overall efficiency. To date, only a handful of SF chromophores have been discovered. This scarcity is commonly recognized as the most significant barrier to realizing SF-based solar cells~\cite{Casanova2018, padula2019singlet, Pradhan_Zeng2022, liu2022finding, zeng2014seeking}. We outline our vision for how the quantum algorithm can be utilized to accelerate the discovery of SF chromophores by elucidating how dynamical processes, such as singlet fission, triplet separation, and charge transfer rates, can be probed and optimized for using our algorithm. Lastly, we provide resource estimates for relevant systems of interest in~\cref{tab:resources}, showing {that $500$ femtoseconds of evolution under a $6$-state $21$-mode vibronic coupling Hamiltonian describing triplet exciton separation~\cite{pradhan2023triplet} requires $154$ qubits and $3.54 \times 10^7$ Toffoli gates using empirically extrapolated Trotter error.}\\% \st{It is worth noting that estimates based on commutator bounds tend to pessimistically overestimate runtime. Hence, actual runtimes will likely be even smaller.}}\\

The manuscript is structured as follows. In~\cref{sec:background}, we briefly review the construction of the vibronic Hamiltonian. We then introduce the quantum algorithm for performing vibronic dynamics, initial state preparation, and extracting observables in Section~\ref{sec:algorithm}. In Section \ref{sec:application}, we outline a proof-of-principle workflow where scalable quantum simulation of non-adiabatic dynamics could provide utility in the context of singlet fission-based solar cell materials discovery.

\section{Background} \label{sec:background}
Vibronic interactions become important whenever electronic states are energetically degenerate or pseudo-degenerate. In the adiabatic basis, the eigenstates of the electronic Hamiltonian interact with one another through non-adiabatic couplings, which can give rise to singularities. A practical workaround for this problem is propagating the molecular time-dependent Schr\"{o}dinger equation using a diabatic electronic basis, in which the couplings are potential terms varying smoothly with nuclear coordinates. For many photochemical scenarios, it is standard procedure to employ the Köppel-Domcke-Cederbaum (KDC) vibronic Hamiltonian~\cite{KDC1984},
\begin{align} \label{eq:KDC_Hamiltonian}
H & = \mathbb{I}_{\text{el}} \otimes (T_{nuc} + V_0) + \boldsymbol{W}.
\end{align}
Here, $T_{nuc}$ and $V_0$ are, respectively, the vibrational kinetic and potential energy operators obtained from the harmonic approximation to the electronic ground state's potential energy surface \cite{meyer2009multidimensional}
\begin{align}
T_{nuc} = \frac{1}{2} \sum_{r} \omega_r P_r^2, \quad\quad\quad V_0 = \frac{1}{2} \sum_r \omega_r Q_r^2,
\end{align}
where $P_r = -i \partial / \partial Q_r$ and $\omega_r$ are, respectively, the momentum operator and the harmonic frequency associated with mode $r$.
The diabatic potential $\boldsymbol{W} = \boldsymbol{W}^{(0)} + \boldsymbol{W}^{(1)} (\vec{Q}) + \boldsymbol{W}^{(2)} (\vec{Q}) + \hdots,$ is Taylor expanded at a reference nuclear geometry, typically that at equilibrium, in a set of dimensionless normal vibrational mode coordinates $\vec{Q}$. To account for spin-forbidden transitions, such as intersystem crossing, $\boldsymbol{W}$ may be augmented with the inclusion of spin-orbit coupling terms $\boldsymbol{W}^{'} = \boldsymbol{W} + \boldsymbol{H}_{SO}$, making \marktext{$H$} a spin-vibronic Hamiltonian~\cite{eng2015spin, zeng2017diabatization, pradhan2023unified}. Given $N$ (spin-)diabatic electronic states (``diabats"), $\boldsymbol{W}^{'}$ is an $N \times N$ block matrix where each block is of the general form
\begin{align}
%\label{eq:W_ii} W_{ii} (\vec{Q}) & = E_i + \sum_{n} r_n^{(i)} Q_n + \sum_{n,m} \gamma_{nm}^{(i)} Q_n Q_m + \hdots  \\ 
\label{eq:W_ij} \boldsymbol{W}^{'}_{ij}(\vec{Q}) = \lambda^{(i,j)} + \sum_{r} a_{r}^{(i,j)} Q_{r} + \sum_{r r^{'}} b_{r r^{'}}^{(i,j)} Q_r Q_{r^{'}} + \cdots, 
\end{align}
The parameters $\lambda^{(i,j)}$, $a_r^{(i,j)}$, $b_{r r^{'}}^{(i,j)}$, $\hdots$ are coupling constants describing the interaction between the $i$th and $j$th diabats induced by the modes labeled by $r$, $r'$, etc. These values can be obtained from electronic structure calculations and diabatization protocols, for which various approaches exist~\cite{koppel1981theory, nooijen2003first, mahapatra2007effects, faraji2008towards, aranda2021vibronic, stanton2007vibronic, atchity1997determination, ruedenberg1993quantum, zeng2017diabatization, zhang2020novel, Neville_Schuurman2024}. In the presence of spin-orbit coupling, the coupling constants may become complex.\\

The accuracy and complexity of the (spin-)vibronic Hamiltonian is determined by the diabatic states and vibrational modes included, the expansion order of $\boldsymbol{W}^{'}$, and the determined coupling constants. Typically, the matrix elements of \cref{eq:W_ij} are truncated to first order, yielding the linear vibronic coupling (LVC) model~\cite{KDC1984}. Truncation to quadratic vibronic coupling (QVC) is also common, depending on the problem~\cite{Bersuker_2006}. The LVC and QVC models often capture qualitative vibronic interactions and suffice for small amplitude distortions~\cite{Lang_Zeng2018}. However, higher-order expansions are needed for quantitative spectra and dynamics involving longer timescales, large amplitude motions, or highly anharmonic potentials~\cite{viel2004, eisfeld2005, bhattacharyya2014, eisfeld2014, mondal2018, parkes_worth2024, meyer2009multidimensional, brown2021unified}.\\

We utilize a discretization over a grid to implement the operators $Q$ and $P$ in a real-space representation. Ref.~\cite{macridin2018digital} provides a detailed derivation of the procedure along with a comprehensive investigation of the discretization error. In particular, they show that the discretization error decreases exponentially with the number of grid points. We will make use of the same discretization protocol. Working in real space with $K = 2^k$ grid points, we take the eigenvectors of $Q$ to be the $k$-qubit computational basis states with eigenvalues 
\begin{equation}\label{eq:discretization_q}
    Q\ket{x} = \Delta (x- K/2) \ket{x},
\end{equation}
for integer $x \in \{0,1,\hdots K-1 \}$ and $\Delta = \sqrt{2\pi/K}$. The momentum operator $P$ is related to $Q$ via the quantum Fourier transform (QFT) as
\begin{equation}\label{eq:discretization_p}
    P = \text{QFT}^\dag \cdot X_{k-1} \cdot Q \cdot X_{k-1} \cdot \text{QFT},
\end{equation}
where $X_{k-1}$ is an $X$ gate on the most significant qubit.

\section{Quantum Algorithm} \label{sec:algorithm}This section describes our algorithm for time evolution under a vibronic Hamiltonian. One key technical contribution is providing the first trotterization scheme for vibronic Hamiltonians extending beyond two electronic states. This is done using the partitioning in~\cref{eq:fragments} and the corresponding technique to block-diagonalize each fragment of the Hamiltonian. Our next key contribution is the heavily-optimized implementation of time evolution for each fragment after block-diagonalization. Namely, we can reduce the cost of exponentiating the block-diagonal matrix in~\cref{eq:action} to essentially have the same cost as the exponentiation of a single block. Lastly, we provide further optimization via a caching protocol described in \cref{sec:Caching} that strategically stores parts of the calculation to avoid redundant computations later.\\

Given $N= 2^n$ electronic states and $M$ vibrational modes, with each mode discretized into $K = 2^k$ points, the vibronic Hamiltonian acting on the space of $M\log(K) + \log(N)$ qubits can be written as $H = T + V$ for
\begin{equation} \label{eq:KDC}
    T = \mathbb{I}_{\text{el}} \otimes \sum_{r=0}^{M-1} \frac{\omega_r}{2}P_r^2, \quad\quad\quad V = \left(\marktext{\mathbb{I}_{\text{el}}} \otimes V_0\right) + \boldsymbol{W}^{\prime} = \sum_{i,j=0}^{N-1} \ket{j}\bra{i}\otimes V_{ji},
\end{equation}
where each $V_{ji} = \sum_{|\vec{\boldsymbol{\alpha}}| \leq d}\, c_{\vec{\boldsymbol{\alpha}}}^{(j,i)}\, \boldsymbol{Q}^{\vec{\boldsymbol{\alpha}}} = \lambda^{(j,i)} + \sum_{r=0}^{M-1}    \gamma_{r}^{(j,i)}  Q_{r} + \sum_{r, r^{'}=0}^{M-1}   \beta_{r r^{'}}^{(j,i)}  Q_{r}Q_{r^{'}} + \cdots$ is a $d$-degree multivariate polynomial of position operators. Here, we have used the multi-index notation $ \boldsymbol{Q}^{\vec{\boldsymbol{\alpha}}} = Q_0^{\alpha_0} Q_1^{\alpha_1} \cdots Q_{M-1}^{\alpha_{M-1}}$. As shown later, position and momentum operators in the real-space representation can be exponentiated cheaply on a quantum computer, making the implementation of operators like $e^{itV_{ij}}$ and $e^{itT}$ straightforward. This motivates the use of Trotter product formulas for implementing $e^{itH}$; however, to do this, we need first to decompose $H=\sum_m H_m$ such that each $e^{itH_m}$ can be implemented in a fast forwardable manner. We do this by writing $V = \sum_{m=0}^{N-1}   H_m$ for
\begin{equation}\label{eq:fragments}
     H_m = \sum_{j=0}^{N-1} \ket{j}\bra{m\oplus j} \otimes V_{j, m\oplus j},
\end{equation}
where $m\oplus j = \sum_{s=0}^{n-1} (\tilde m_s \oplus \tilde j_s) \cdot 2^s$ is the integer resulting from the bitwise XOR product between the bitstring representations of integers $m = \sum_{s=0}^{n-1} \tilde m_s \cdot 2^s$ and $j = \sum_{s=0}^{n-1} \tilde j_s \cdot 2^s$. 
Notice that the block-diagonal fragment $e^{itH_0} = e^{i t\sum_j \ket{j}\bra{j}\otimes V_{jj}} = \sum_j \ket{j}\bra{j}\otimes  e^{i tV_{jj}}$ reduces to a sequence of evolutions $e^{itV_{jj}}$ controlled by the corresponding electronic state $\ket{j}$. Our strategy then becomes clear: block-diagonalize each $e^{itH_m}$ and implement them like $e^{itH_0}$. For simplicity, in what follows, we assume the coefficients in $V_{ji}$ are real, implying $V_{ji} = V_{ij}$. This assumption is not essential as the algorithm described below can be adapted to the case where $V_{ji} = V_{ij}^\dag$ by separating the Hermitian and anti-Hermitian parts of $V_{ji}$.\\

\begin{figure}[h]
  \centering
  \includegraphics[width=0.75\linewidth]{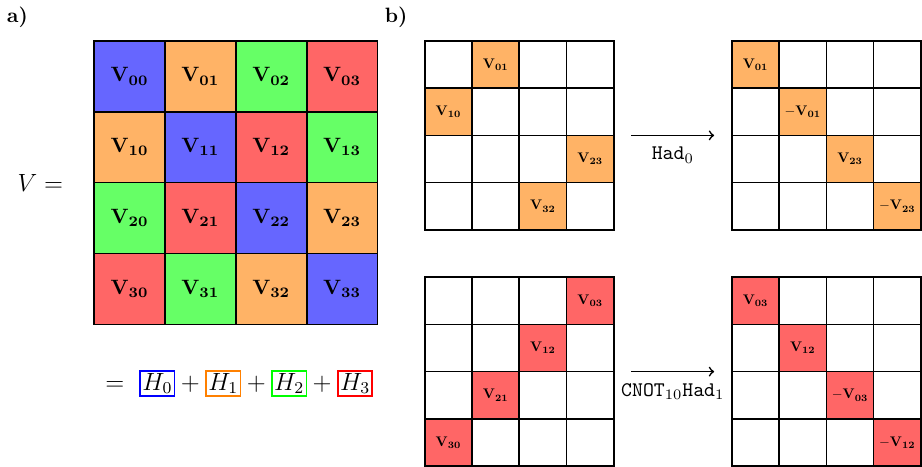}
  \caption{{a) Fragmentation of the vibronic potential $V$ for the $N=4$ case. b) Illustrative examples of the Clifford block-diagonalization for fragments $H_1$ and $H_3$. $\texttt{Had}_i$ denotes application of Hadamard gate on qubit $i$, and $\texttt{CNOT}_{ij}$ denotes a CNOT controlled by qubit $i$ and targetting qubit $j$.}}
  \label{fig:trotterscheme}
\end{figure}

If $\tilde m$ has Hamming weight 1, that is $j$ and $m\oplus j$'s bitstring representations only differ in a single bit, we can block-diagonalize $H_m$ by applying \marktext{the Hadamard gate \texttt{Had}} on both sides of the qubit where they differ since $(\texttt{Had}) X (\texttt{Had}) = Z$. To see this, without loss of generality, assume the rightmost qubit is the differing one; then, we have
\begin{equation}
      (\texttt{Had})H_m(\texttt{Had}) = \sum_{j'= 0}^{N/2-1} \ket{j'}\bra{j'} \otimes (\texttt{Had}) X (\texttt{Had}) \otimes V_{(2j'),(2j'+1)} = \sum_{j'= 0}^{N/2-1} \ket{j'}\bra{j'} \otimes Z \otimes V_{(2j'),(2j'+1)}.
\end{equation}
If $\tilde m$ has Hamming weight greater than 1, we construct a unitary $U$ by choosing one of the qubits where $\tilde m$ is 1 and using it as a control to apply a \texttt{CNOT} to all other qubits where $\tilde m$ is 1. This ensures that the pairs $U\ket{j}\bra{m\oplus j}U^\dag$ only differ in a single qubit, the qubit used as the control. To see this, notice how a \texttt{CNOT} gate fixes one of the mismatchings between two pairs differing in both bits: 
\begin{equation}
    \mathtt{CNOT} \ket{b_2 b_1\vphantom{\overline b_2 \overline b_1}}\bra{\overline b_2 \overline b_1} \mathtt{CNOT} = \ket{b_1\oplus b_2, b_1 \vphantom{\overline b_2 \overline b_1}}\bra{\overline b_1 \oplus \overline b_2, \overline b_1} = \ket{b_1\oplus b_2, b_1 \vphantom{\overline b_2 \overline b_1}}\bra{ b_1 \oplus b_2, \overline b_1},
\end{equation}
where we've used $b_1\oplus b_2 = \overline b_1 \oplus \overline b_2$. Hence, $U H_m U^\dag$ can be block-diagonalized by applying a Hadamard on the control qubit. {A visual representation of the fragmentation scheme and examples of the Clifford block-diagonalization are provided in Figure \ref{fig:trotterscheme}}.
We have shown how to block-diagonalize each fragment $e^{itH_m}$ to be implemented in the same manner as $e^{it H_0}$, now we give our algorithm for the implementation of $e^{i tH_0}$. For simplicity, we absorb $t$ into $H_0$ for the rest of the discussion.\\

A naive implementation of $e^{iH_0} = e^{i \sum_j \ket{j}\bra{j}\otimes V_{jj}} = \sum_j \ket{j}\bra{j}\otimes  e^{i V_{jj}}$ would be to perform the corresponding evolution $e^{iV_{jj}}$ on the vibrational space controlled by the corresponding state $\ket{j}$ for all $N$ electronic states. This would have a cost of $N$ times that of a single controlled evolution $e^{iV_{jj}}$. Our next key result is an algorithm to implement $e^{i H_0}$ with essentially the same cost as implementing a single uncontrolled evolution $e^{iV_{jj}}$. From \cref{eq:discretization_q}, we have that $Q\ket{x} = \Delta (x- 2^{k-1}) \ket{x}$. However, if we interpret the computational basis state bitstrings in the signed integer representation, we will have $Q\ket{x} = \Delta\cdot x \ket{x}$. Accordingly, all the arithmetic operations below will correspond to signed arithmetic. Using this framework, we can write the action of $e^{iH_0}$ on any computational basis state $\ket{j}\ket{\boldsymbol{x}} = \ket{j}\bigotimes_{r=0}^{M-1} \ket{x_r}$ as
\begin{align}\label{eq:action}
    \sum_j \ket{j}\bra{j}\otimes  e^{i V_{jj}}  &= \sum_{j, \boldsymbol{x}} \ket{j}\bra{j} \otimes \ket{\boldsymbol{x}}\bra{\boldsymbol{x}} \cdot \exp\left(i\sum_{|\vec{\boldsymbol{\alpha}}| \leq d}\, \Delta^{|\vec{\boldsymbol{\alpha}}|}\cdot c_{\vec{\boldsymbol{\alpha}}}^{(j,j)}\, \boldsymbol{x}^{\vec{\boldsymbol{\alpha}}}\right)\nonumber \\ &= \prod_{|\vec{\boldsymbol{\alpha}}| \leq d} \sum_{j,\boldsymbol{x}} \ket{j}\bra{j} \otimes \ket{\boldsymbol{x}}\bra{\boldsymbol{x}} e^{i \Delta^{|\vec{\boldsymbol{\alpha}}|}\cdot c_{\vec{\boldsymbol{\alpha}}}^{(j,j)}\, \boldsymbol{x}^{\vec{\boldsymbol{\alpha}}}}.
\end{align}
Therefore, we can implement $e^{iH_0}=\sum_j \ket{j}\bra{j}\otimes  e^{i V_{jj}}$ by implementing each $\sum_{j,\boldsymbol{x}} \ket{j}\bra{j} \otimes \ket{\boldsymbol{x}}\bra{\boldsymbol{x}} \cdot\exp\left(i \Delta^{|\vec{\boldsymbol{\alpha}}|}\cdot c_{\vec{\boldsymbol{\alpha}}}^{(j,j)}\, \boldsymbol{x}^{\vec{\boldsymbol{\alpha}}}\right)$ in \cref{eq:action}. To implement each $\sum_{j,\boldsymbol{x}} \ket{j}\bra{j} \otimes \ket{\boldsymbol{x}}\bra{\boldsymbol{x}} \cdot\exp\left(i \Delta^{|\vec{\boldsymbol{\alpha}}|}\cdot c_{\vec{\boldsymbol{\alpha}}}^{(j,j)}\, \boldsymbol{x}^{\vec{\boldsymbol{\alpha}}}\right)$ we
\begin{enumerate}\label{eq:steps}
    \item Load the binary representation of the coefficients {$\Delta^{|\vec{\boldsymbol{\alpha}}|}\cdot c_{\vec{\boldsymbol{\alpha}}}^{(j,j)}$} in an ancillary register controlled on the corresponding electronic state using a series of multi-controlled multi-$X$ gates.
    \item Compute the corresponding variable $\boldsymbol{x}^{\vec{\boldsymbol{\alpha}}}$ in another register using signed quantum arithmetic by taking the product between the corresponding mode registers.
    \item {Compute the product between the coefficient and the variable registers $\Delta^{|\vec{\boldsymbol{\alpha}}|} \cdot c_{\vec{\boldsymbol{\alpha}}}^{(j,j)}\, \boldsymbol{x}^{\vec{\boldsymbol{\alpha}}}$ in an ancilla register.
    \item Perform a phase gradient operation on the register holding the value $\Delta^{|\vec{\boldsymbol{\alpha}}|}\cdot c_{\vec{\boldsymbol{\alpha}}}^{(j,j)}\, \boldsymbol{x}^{\vec{\boldsymbol{\alpha}}}$.}
    \item Uncompute all intermediate results.
\end{enumerate}
Here a $b$-qubit phase gradient operation is defined as $\sum_{y=0}^{2^b -1} e^{i 2\pi y/2^b} \ket{y}\bra{y}$, { and is performed via modular addition to a register holding the resource state $\ket{R} = \frac{1}{2^{b/2}}\sum e^{i 2\pi y/2^b} \ket{y}$. Such a resource state only needs to be prepared once at the beginning of the algorithm and is re-used throughout the entire computation. Another advantage of implementing the phase gradient this way is that it allows one to skip step 3 in the above and directly add the product of the 2 registers to the resource register as shown in \cref{fig:poly_evolution}.}
\begin{figure}[t!]
    \centering
        \begin{quantikz} 
            \lstick{$\ket{j}$} &\qwbundle{\log N} && \gate{j}\wire[d][4]{q} &&&&  \gate{j}\wire[d][4]{q} && \\
            \lstick{$\ket{x_0}$} & \qwbundle{\log K} &&& \gate[3]{\hspace{0.3cm}\mathtt{Mult}\hspace{0.3cm}} \gateinput{$a$} \gateoutput{$a$} && \gate[3]{\hspace{0.3cm}\mathtt{Mult}^\dagger\hspace{0.3cm}} \gateinput{$a$} \gateoutput{$a$} &&& \\ 
            \lstick{$\ket{x_1}$} &\qwbundle{\log K} &&& \gateinput{$b$} \gateoutput{$b$} && \gateinput{$b$} \gateoutput{$b$} &&& \\
            \lstick{$\ket{\vec 0}$} &\qwbundle{2\log K} &&& \gateinput{$0$} \gateoutput{$ab$} & \gate[3]{\hspace{0.3cm}\mathtt{Mult}\hspace{0.3cm}}\gateinput{$a$} \gateoutput{$a$} &\gateinput{$ab$} \gateoutput{$0$}&&&\\
            \lstick{$\ket{\vec 0}$} & \qwbundle{\log 1/\delta} && \gate{\Delta^2c_{\left(1,1\right)}^{\left(j,j\right)}} && \gateinput{$b$} \gateoutput{$b$} &&  \gate{\Delta^2c_{\left(1,1\right)}^{\left(j,j\right)}} &&\\
            \lstick{$\ket{R}$} &\qwbundle{\log{1/\delta}} &&&&\gateinput{$c$} \gateoutput{$c+ ab$}&&&&
        \end{quantikz}
    \caption{{Example circuit implementation for a term of the form $\sum_j \ket{j}\bra{j}\otimes  \exp\left(i \Delta^2 c_{(1,1)}^{(j,j)} x_0x_1\right)$. Where $\delta$ is the fixed point precision used to represent the coefficients in the Hamiltonian, \marktext{and \texttt{Mult} denotes a quantum arithmetic multiplication gate as described in Ref.~\cite{su2021fault}.}}}\label{fig:poly_evolution}
\end{figure}

Lastly, we show the implementation of $e^{iT}$. The operator $e^{iT}$ reduces to $\exp{\left(\frac{i}{2}\sum_{r=0}^{M-1} \omega_r P_r^2\right)}$ on the vibrational space. It can be implemented by applying $\exp{\left(\frac{i}{2}\omega_r P_r^2\right)}$ to each corresponding mode, where $P_r$ is the momentum operator of the $r^{th}$ mode related to $Q_r$ by \cref{eq:discretization_p}. Hence, after diagonalizing via QFT, $\exp{\left(\frac{i}{2}\omega_r P_r^2\right)}$ can be implemented in a similar manner as $\exp{\left(\frac{i}{2}\omega_r Q_r^2\right)}$. Writing $H_N = T$ for completeness, we can have $H = \sum_{m=0}^{N} H_m$. Hence, we can implement $e^{itH}$ using a Trotter product formula. The first-order formula $U_1(\theta)$ and second-order formulas $U_2(\theta)$ will have the form
\begin{align}\label{eq:Trotter}
    U_1(\theta) &= \prod_{m=0}^N e^{i \theta H_m} = e^{i\theta H} + \mathcal{O}(\theta^2),\\
    U_2(\theta) &= \prod_{m=0}^N e^{i \theta H_m} \prod_{m=N}^0 e^{i \theta H_m}  = e^{i\theta H} + \mathcal{O}(\theta^3). \label{eq:Trotter2nd}
\end{align}

\subsection{Optimization via Caching}\label{sec:Caching}

In this section, we provide a further optimization of the algorithm based on a caching protocol. The motivation behind this approach is to avoid redundant calculations when implementing higher-degree terms in the polynomial. Consider the implementation of a term like \(\exp\left(i \theta\, Q_{s} Q_{t} Q_{r}\right)\) in our algorithm. This requires computing the product of the corresponding modes \(\ket{x_s x_t x_r}\) which involves first calculating \(\ket{x_s x_t}\) and then multiplying the result with \(\ket{x_r}\) to obtain \(\ket{x_s x_t x_r}\). However, we have already computed the intermediate product \(\ket{x_s x_t}\) when implementing the term \(\exp\left(i \theta' \, Q_{s} Q_{t}\right)\). Instead of uncomputing this intermediate result after its use, we can cache it for reuse. By retaining \(\ket{x_s x_t}\), we can compute \(\ket{x_s x_t x_r}\) by a single multiplication with \(\ket{x_r}\), thereby saving an additional computation and uncomputation of $\ket{x_sx_t}$. In fact, we can do this for every $r=0,..., M-1$. More generally, we can extend this idea to terms of arbitrary degree. Terms of degree \(L\) can reuse computations from terms of degree \(L-1\), which in turn reuse computations from terms of degree \(L-2\), and so on.\\

For the most space-efficient implementation of this caching protocol, we utilize a \textit{depth-first search} approach in determining the order in which we exponentiate the terms in the polynomial. We organize the computation of terms in a tree-like structure where each node represents a term, and edges represent the dependency of higher-degree terms on lower-degree ones. Then, the order of implementation is given by a depth-first search traversal of this tree starting from terms of degree 1. This ensures that every term of degree \(L > 1\) reuses the computation from a term of degree \(L - 1\) in a way that, at any given time, we are only caching one term at each degree.

% \begin{figure}[t!]
%     \centering
%     \includegraphics[width=0.9\textwidth]{Multadd.png}
%     \vspace{-2.8cm}
%     \caption{{Implementation of the $\mathtt{Mult\,\, Add}$ operation that performs the modular addition of the product of 2 registers of size $k$ and $n$ onto a third register using a sequence of controlled modular additions.}}
%     \label{fig:mult_add}
% \end{figure}
\subsection{Complexity Analysis}

Here, we present the runtime and space complexity of our algorithm for implementing \marktext{time evolution using a general order Trotter product formula}. The proof and a more detailed analysis are provided in \cref{ap:proof}. 
\marktext{
\begin{theorem}\label{thm:complexity}
    Given an $N$-state, $M$-mode KDC Hamiltonian $H = T + V$ with $T$ and $V$ as defined in \cref{eq:KDC} expanded up to $d^{th}$ order with each mode discretized into $K$ gridpoints, our algorithm presented in this section can implement $e^{iHt}$ on the system of $M\log(K)+\log(N)$ qubits via a $p^{th}$ order Trotter formula using 
    \[
    \#\, T\, \text{gates}\in O(5^{p/2}\cdot t^{1+1/p}\cdot \epsilon^{-1/p}\cdot \Omega),
    \]
    where $\Omega$ is the cost of a single first order Trotter step,
    \[
    \Omega\in O(M^d(N\cdot d\cdot\log^2(K) + N^2 )),
    \]
    which requires $O(d^2\cdot \log(K)+\log(N))$ ancilla qubits.
\end{theorem}$ $\\
The scaling with respect to error $\epsilon$ and product formula order $p$ comes from standard results in the theory of Trotter error~\cite{papageorgiou2012efficiency, childs2021theory}. The specific complexity of our algorithm is thus encapsulated in the factor $\Omega$, which is highly dependent on the degree of the expansion in $V_{ji}$, since a $d$-degree multivariate polynomial over $M$ modes will generally have $O(M^d)$ terms and we perform time evolution for each term at least once. In practice $d$ is small, often around $2$ to $4$, and higher degree terms are sparse, usually restricted to terms involving no more than two different modes at a time, e.g., $Q_{r}^2Q_{r^{'}}$. The number of terms is then multiplied by the cost of time evolution for each term, which is determined by cost of loading coefficients for $N$ electronic states and performing arithmetic operations on registers of $O(\log K)$ qubits up to degree $d$. We obtain linear instead of quadratic dependence in $d$ due to caching. 
}

% Notice that the complexity of our algorithm is highly dependent on the degree of the expansion in $V_{ji}$ since a $d$-degree multivariate polynomial over $M$ modes will generally have $\mathcal{O}(M^d)$ terms. In practice, $d$ is small, often around $2$ to $4$, and higher degree terms are sparse, usually restricted to terms involving no more than $2$ different modes at a time (e.g., $Q_{r}^2Q_{r^{'}}$).

\subsection{Initial State Preparation} \label{sec:state_prep}

Preparing physically relevant initial states is crucial in extracting useful information from simulations. For photo-induced dynamics, the initial state often corresponds to a vertical excitation of the system, which has a simple product form 
\begin{equation}\label{eq:initial_state1}
    \ket{\psi (0)} = \ket{j}_{\text{el}} \bigotimes_{r=0}^{M-1} \ket{\chi_0},
\end{equation}
where $\ket{j}_{\text{el}}$ is the corresponding excited electronic state (represented as a computational basis in our model), and $\ket{\chi_0}$ is the harmonic oscillator ground state. State $\ket{\chi_0}$ corresponds to the discretized Hermite-Gauss function of zeroth order. If each mode is discretized into $K = 2^k$ grid points, $\ket{\chi_0}$ is then the $k$-qubit state
\begin{equation}
    \ket{\chi_0} = \frac{1}{Z} \sum_{x=0}^{K-1} \exp\left(\frac{-\pi\cdot\left(x-\frac{K}{2}\right)^2}{K}\right) \ket{x},
\end{equation}
where $Z$ is the normalization constant such that $\|\ket{\chi_0}\|=1$. Methods to prepare Gaussian states are discussed in Refs. \cite{kitaev2008wavefunction, grover2002creating,iaconis2024quantum}. Instead of starting in a specific electronic state, we can choose to start in a superposition of electronic states. This choice is especially relevant for absorption/emission spectroscopy, where the initial state of the electronic register is the superposition resulting from the electronic ground state being acted upon by the dipole operator $\mu$:
\begin{align}\label{eq:initial_state2}
    \ket{\psi (0)} &= \frac{\mu\ket{0}_{\text{el}}}{\|\mu\ket{0}\|} \bigotimes_{r=0}^{M-1} \ket{\chi_0} = \left(\sum_{j=0}^{N-1} \mu_{j,0}\, \ket{j}\right) \bigotimes_{r=0}^{M-1} \ket{\chi_0},
\end{align}
where $\mu_{j,0} = \frac{\bra{j}\mu\ket{0}}{\|\mu\ket{0}\|}$. $\sum_{j=0}^{N-1} \mu_{j,0}\, \ket{j}$ is a state of small dimensionality that can prepared using any black-box state preparation algorithm \cite{gosset2024quantumstatepreparationoptimal, sanders2019black, wang2021fast}. The product form of the states in \cref{eq:initial_state1} and  \cref{eq:initial_state2} together with the fact that the size of each register will be small in practice, the cost of state preparation will be negligible compared to the cost of the simulation.
\subsection{Observables} \label{sec:observables}

Following the development and analysis of our algorithm, the next critical step is identifying relevant observables for extracting meaningful information from simulations. Key observables of interest in vibronic simulations include electronic state populations and spectroscopic quantities.

\subsubsection{Electronic State Populations}

In the context of vibronic simulations, electronic state populations describe the probability of finding the system in a particular diabatic electronic state at a given time. For $\ket{\psi(t)} = e^{i Ht} \ket{\psi(0)}$, the population of diabatic state $\ket{j}$ at time $t$ is given by
\begin{align} \label{eq:population}
p_j(t) = \bra{\psi(t)}\left( \ket{j}\bra{j} \otimes \mathbb{I}_{\text{vib}}\right)\ket{\psi(t)},
\end{align}
\marktext{where $\mathbb{I}_{\text{vib}}$ denotes the identity on the vibrational mode registers.} In our model, estimating Eq.(\ref{eq:population}) simply corresponds to measuring the electronic register in the computational basis and collecting statistics. This allows us to simultaneously estimate populations of all states using $\mathcal{O}(\epsilon^{-2})$ measurements for an $\epsilon$ accurate estimation. Alternatively, to estimate the population of any particular state, the quantum amplitude estimation (QAE) algorithm \cite{qae} can be employed, which, at the cost of a longer circuit depth, reduces the total runtime to $\mathcal{O}(\epsilon^{-1})$.\\

The state populations obtained throughout propagation can then be straightforwardly fitted to exponential functions or applied within more sophisticated kinetic models to extract transition rates and excited-state lifetimes \cite{marquetand2016challenges}. This ability enables the study of key physical phenomena such as non-radiative relaxation via intersystem crossing and internal conversion, as well as exciton transfer and charge transfer. Studying these phenomena is crucial for uncovering reaction pathways in photochemistry, as well as designing materials for optoelectronic and photovoltaic technologies.

\subsubsection{Spectroscopic Quantities}

Spectroscopic quantities are another class of fundamental observables in vibronic simulations. Specifically, absorption and emission spectra are readily obtained via a Fourier transform of the dipole autocorrelation function
\begin{align}
    C(\omega) &= \frac{1}{2\pi}\int_{-\infty}^\infty dt\, e^{-i\omega t}\bra{\psi} e^{i H t} \mu e^{-i H t} \mu\ket{\psi},
\end{align}
which provide insights into the interaction of molecules with electromagnetic radiation. Here, ${\mu}$ is the dipole moment operator, and $\ket{\psi}$ is the reference state,  often taken to be the state from \cref{eq:initial_state2}.\\

The recently developed framework of generalized quantum phase estimation and related techniques~\cite{loaiza2024nonlinear, kharazi2024efficient} enable the efficient extraction of linear and non-linear spectroscopic quantities on a quantum computer. In particular, the technique of generalized quantum phase estimation~\cite{loaiza2024nonlinear} allows one to directly sample from the absorption spectra of a system with an energy resolution $\eta$ when given access to the time evolution of the system for $t \sim \mathcal{O}(1/\eta)$. The absorption and emission spectra are essential for understanding and predicting optical properties such as brightness, color purity, and stability in materials; this can guide the design of new materials for optoelectronic devices like organic lasers and light-emitting diodes (OLEDs).

\section{Applications in singlet fission solar cell design} \label{sec:application}
This section discusses how our quantum algorithm from \cref{sec:algorithm} could be integrated into a materials discovery pipeline in an industrial setting. While simulations of excited-state dynamics have extensive industrial applications ranging from optoelectronic technologies to non-invasive cancer therapies~\cite{dougherty1998photodynamic}, here we focus on the specific case of singlet fission (SF) solar cells and how our algorithm can be utilized to accelerate the discovery of candidate SF chromophores.\\

% \subsection{Overview}

As global energy demands rise, the reliance on unsustainable energy sources poses significant environmental and economic challenges. Solar energy offers a scalable and sustainable solution; however, conventional photovoltaics are restricted by the Shockley-Queisser limit~\cite{ShockleyQueisser1961}, which puts an upper bound of $\sim 33 \%$ on the power conversion efficiency that a single junction solar cell can have. Consequently, significant efforts have been put towards going beyond this limit with \textit{singlet fission} (SF)~\cite{Baldacchino2022}, a type of multi-exciton generation process, emerging as a promising direction. SF-based solar cells can, in principle, achieve a theoretical internal quantum efficiency of $200 \%$: two electrons for one photon \cite{SM2010, Casanova2018, Japahuge_Zeng2018, FG2019}. However, commercialization of SF solar cells has been hindered by the fact that only a handful of candidate SF chromophores have been discovered to date. This scarcity is commonly recognized as the most significant barrier preventing the realization of SF technologies~\cite{Casanova2018, padula2019singlet, Pradhan_Zeng2022, liu2022finding, zeng2014seeking}.\\

To overcome this challenge, computational search via \textit{ab initio} simulations has been proposed as a promising route to accelerating the discovery of viable SF-based solar cell materials~\cite{Japahuge_Zeng2018}. Energetic criteria can serve as a good initial screening tool as they provide insight into whether the process of SF is energetically allowed. However, the rate of SF is of greater importance to the success of an SF chromophore, as the process must outcompete other relaxation paths operating on ultrafast timescales. Despite this, proposals so far for the computational design and screening of new SF chromophores have almost entirely relied on assessment based on static electronic energy criteria ~\cite{padula2019singlet, weber2022machine, veilleux2024designing, wang2024computational}. 
%This is because SF and other important downstream processes leading to the generation of free-charge carriers are facilitated via vibronic interactions whose accurate modeling requires a fully quantum description of the non-adiabatic dynamics
However, considering only electronic excitation energies can be far from being enough if a quantitative description of the SF process is desired. This is because SF is an excited-state dynamical process in which vibronic interaction is of central importance and should not be ignored~\cite{smith2013recent, Xie_Liu2019, tempelaar2017vibronic, teichen2012microscopic, Berkelbach2013_I, berkelbach2013microscopicII, bakulin2016real, tamura2015first, yao2016coherent, fujihashi2017effect, morrison2017evidence, miyata2017coherent, Japahuge_Zeng2018, Casanova2018, casillas2020molecular}. Semi-classical methods such as surface hopping fail to capture important quantum effects involved in SF, such as tunneling \cite{chan2013quantum, parenti2023quantum}. On the other hand, tackling the fully quantum simulation via methods like MCTDH beyond a model with reduced dimensionality quickly becomes too resource-intensive, or intractable altogether, for integration into a computational discovery pipeline{, as discussed in more detail in Appendix \ref{app:classical_hardness}}. A highly optimized and scalable quantum algorithm like the one provided in \cref{sec:algorithm} lifts the above bottleneck and allows for a more effective computational search of new SF chromophores.

\subsection{Overview}

To provide a concrete example of how our quantum algorithm for non-adiabatic dynamics could enable the discovery of new SF materials for solar cells, a high-level discovery workflow is illustrated in~\cref{fig:workflow}. Starting with an initial candidate molecule, a vibronic model Hamiltonian is constructed from electronic structure calculations, utilizing either classical or quantum resources. Once the vibronic Hamiltonian has been constructed, our algorithm performs the subsequent non-adiabatic dynamics simulations from where it extracts relevant observables such as state populations or absorption spectra as described in Section \ref{sec:observables}. This dynamical information is then used to guide modifications (e.g., functional group or heteroatom substitutions~\cite{chen2014effects, zeng2014seeking, bhattacharyya2015small}) to the candidate molecule, either through expert interpretation or machine learning approaches such as Bayesian optimization~\cite{hernandez2017parallel, pyzer2018bayesian, griffiths2020constrained}. This process could be iterated to optimize for properties of interest or generate high-quality data for complimentary computer-aided design workflows. Furthermore, the ability to rapidly perform non-adiabatic simulations following the modification of a molecular system can provide indispensable mechanistic insights into properties of interest. For instance, changes to the state population dynamics following molecular modification would offer valuable information on the complicated interplay between molecular geometry, electronic structure, and dynamical pathways, from which revised rational design principles could be proposed~\cite{ito2018molecular, hasobe2021molecular, wang2021molecular}.\\

\begin{figure}[t!]
    \centering
    \includegraphics[width=0.9\linewidth]{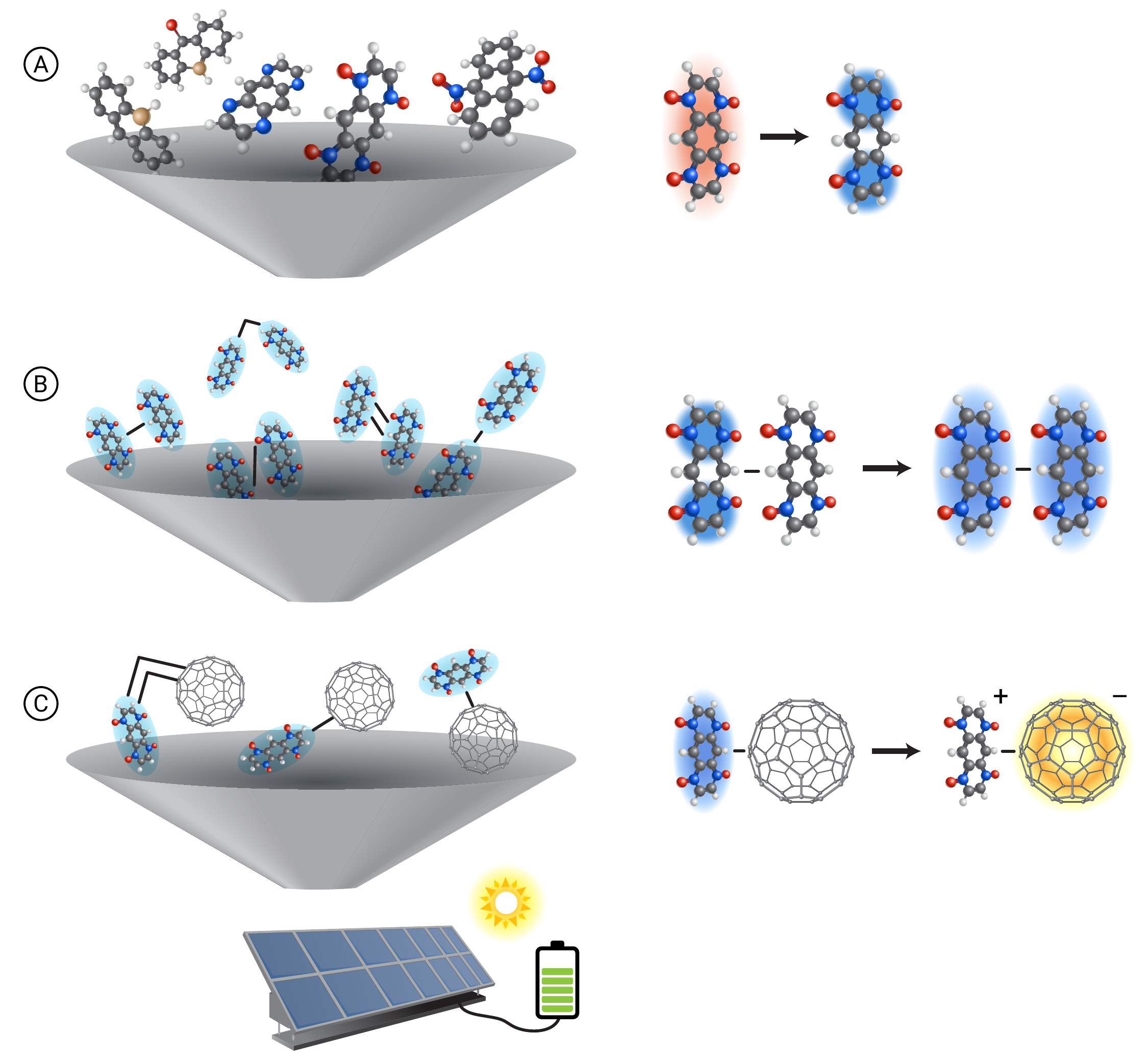}
    \caption{A proof-of-principle SF chromophore design pipeline utilizing the iterative workflow from \cref{fig:workflow}, with stages for (A) singlet fission, (B) triplet separation, and (C) charge transfer optimization. (A): A set of candidate SF chromophores is assessed based on the theoretical rate of singlet fission. Depicted is a set of anthracene derivatives generated via functional group and heteroatom substitutions (left) and a pictorial description of the SF process (right), where a singlet exciton state (red) converts to two triplets (blue) localized at the ends of the chromophore. (B): Following the selection of an SF chromophore, covalent linking strategies are optimized to enable efficient triplet separation. Various bridges and conformational arrangements would be assessed (left) to provide the most rapid triplet separation, where a triplet-pair local to a single SF chromophore converts to two triplets isolated to individual chromophores (right). (C): Finally, the charge transfer rate could be optimized using different chromophore/acceptor strategies, such as chromophore/acceptor covalent linking (left), in analogy to the optimization of step (B), to ensure efficient transfer from a triplet exciton to a charge-separated state at the acceptor interface (right).}
    \label{fig:funnels}
\end{figure}

In the following subsection, we describe a more detailed proof-of-principle pipeline for the design of new SF chromophores as illustrated in \cref{fig:funnels} by applying the workflow of \cref{fig:workflow} to the various vibronically driven processes crucial in determining the efficiency of SF solar cells. We briefly introduce each process, their modeling on a quantum computer, and the corresponding restricted space over which the computational search could be performed. Furthermore, we report \marktext{empirically estimated quantum resource requirements when using a second order product formula \cref{eq:Trotter2nd}} for relevant systems in \cref{tab:resources} to demonstrate the efficiency of our quantum algorithm.\\

\subsection{Design pipeline} \label{sec:design_pipeline}

\begin{figure}[t!]
    \centering
    \includegraphics[width=0.7\linewidth]{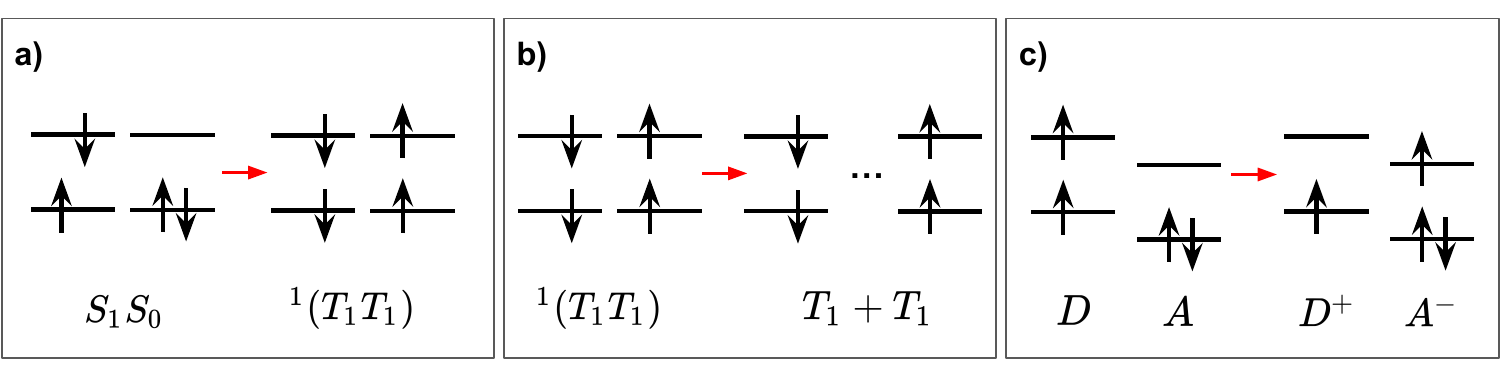}
    \caption{\marktext{A simplified orbital depiction of the processes relevant to the performance of a SF chromophore layer, including a) the fission of a photoexcited singlet into two adjacent triplet excitons, b) separation of the spin-coupled triplet pair, and c) charge transfer from an excitonic donor to an acceptor at the chromophore-acceptor interface.}}
    \label{fig:sf_processes}
\end{figure}

Achieving a high internal quantum efficiency in an SF-based photovoltaic cell depends not only on SF taking place but also on various subsequent processes prior to the generation of free-charge carriers. For one, following SF, triplets must separate so they can migrate to the chromophore-acceptor interface of the photovoltaic cell. Finally, migrated triplets must efficiently undergo a charge transfer with an acceptor at this interface to produce a pair of free charge carriers. Not only is the SF process itself non-adiabatic, but triplet separation and chromophore-acceptor charge transfer typically involve nuclear motion as well, as their portrayal involves conical intersections or avoided crossings ~\cite{prezhdo2009photoinduced, sissa2014vibrational, pradhan2023triplet}, making their dynamics difficult to describe classically. To date, a severely limited number of works have addressed atomistic simulation of the non-adiabatic dynamics involved in triplet separation ~\cite{pradhan2023triplet} and chromophore-acceptor charge transfer ~\cite{xie2015full}. This is inadequate since all three non-adiabatic processes are important in an SF-promoted photovoltaic cell. Ideally, the photon-to-exciton generation efficiency is doubled, the triplet pairs readily separate, and each triplet converts to a pair of charge carriers. These transport processes are schematically depicted in Figure \cref{fig:sf_processes} In the following, we describe each of these processes in further detail and propose an SF chromophore/acceptor design pipeline illustrated in \cref{fig:funnels}. We outline promising computational search spaces for each characteristic of interest to apply the quantum-enabled discovery workflow of \cref{fig:workflow} to each of the dynamical processes discussed.

\vspace{-0.2cm}
\subsubsection{Singlet fission}\label{sec:sf}

\begin{table}[t!]
    \centering
    \renewcommand{\arraystretch}{1.5} % Adds vertical space
    \setlength{\tabcolsep}{12pt} % Adds horizontal space
    \begin{tabular}{c|c|c|c}
        System & \# Qubits & \# Toffoli Gates & Parameters\\ \hline\hline
        \multirow{2}{*}{\vspace{-0.3cm}\textbf{(NO)$_4$-Anth}~\cite{Pradhan_Zeng2022}} 
            & \multirow{2}{*}{{\vspace{-0.3cm}$146$}} 
            & {$5.47 \times 10^6$} 
            & \makecell{$N=5, M=19,$ \\ $t = 100 \text{fs}, \epsilon = 10\%$} \\ \cline{3-4}
            & 
            & {$1.73 \times 10^7$} 
            & \makecell{$N=5, M=19,$ \\ $t = 100 \text{fs}, \epsilon = 1\%$} \\ \hline
        \multirow{2}{*}{\vspace{-0.3cm}\textbf{(NO)$_4$-Anth Dimer}~\cite{pradhan2023triplet}} 
            & \multirow{2}{*}{{\vspace{-0.3cm}$154$}} 
            & {$2.76 \times 10^6$} 
            & \makecell{$N=6, M=21,$ \\ $t = 100 \text{fs}, \epsilon = 1\%$} \\ \cline{3-4}
            & 
            & {$3.54 \times 10^{7}$} 
            & \makecell{$N=6, M=21,$ \\ $t = 500 \text{fs}, \epsilon = 1\%$} \\ \hline
        \multirow{2}{*}{\vspace{-0.3cm}\textbf{Anth/C$_{60}$}~\cite{xie2015full}} 
            & {$113$} 
            & {$6.62 \times 10^5$} 
            & \makecell{$N=4, M=11,$ \\ $t = 100 \text{fs}, \epsilon = 1\%$} \\ \cline{2-4}
            & {$1053$} 
            & {$2.66 \times 10^7$} 
            & \makecell{$N=4, M=246,$ \\ $t = 100 \text{fs}, \epsilon = 1\%$}
    \end{tabular}

    \caption{Estimated implementation costs of our quantum algorithm on various systems of interest \marktext{using second order product formula \cref{eq:Trotter2nd}}. The model on top is a $N=5$ state $M=19$ mode quadratic vibronic coupling (QVC) singlet fission model analyzed at $10\%$ and $1\%$ error tolerances, representing decision vs. quantitative problem requirements as per \cref{sec:sf}. Next, a QVC triplet separation model with $N=6$ states and $M=21$ modes is analyzed for time propagations of $t=100$ femtoseconds and $t=500$ femtoseconds. Lastly, we analyze a $N=4$ state linear coupling (LVC) model of charge transfer in a reduced dimensionality setting of $M=11$ modes and the full dimensionality of $M=246$ modes. For all entries, we discretize each mode into $K=16$ grid points.  {The number of Trotter steps utilized in the resource estimations are obtained through extrapolation of empirical Trotter errors, \marktext{available in Ref ~\cite{VibeCounting}.}}}
    \label{tab:resources}
\end{table}

SF is the process of a photoexcited chromophore in the first excited singlet state, $S_1$, transferring energy to a neighboring chromophore in its ground state $S_0$, generating two \textit{triplet} excitons:
\begin{align} \label{eq:sf}
\ce{
$S_0 S_1$ <=> $^1(T_1 T_1)$.
}
\end{align}
Since the pair of triplets ($T_1$) is coupled as an overall singlet state denoted $^1(T_1 T_1)$, SF is a spin-allowed process. Hence, it can happen on a sub-picosecond timescale and outcompete decay pathways such as fluorescence from $S_1$. One feasible approach to designing improved SF chromophores is through the derivatization of known molecular classes that undergo SF or have the potential to undergo SF~\cite{zhang2024novel, patra2024designing}. Derivatization involves the chemical modification of a molecule to create derivatives with altered properties, often including the introduction or modification of functional groups. For instance, a chromophore with theoretically record-breaking fast SF was proposed by introducing four N-oxyl (NO) radical fragments to anthracene~\cite{Pradhan_Zeng2022}, which we call (NO)$_4$-Anth. A challenge faced by the functional group modulation of SF chromophores is searching through the space of possible functional groups and their specific positionings on the parent molecule. We hope this challenge can be overcome via fast and efficient quantum dynamics simulations on a quantum computer.\\

%Simulation in Pipeline:
Following the substitution of a functional group of a candidate molecule, a vibronic Hamiltonian that includes the relevant states must be generated to perform the corresponding dynamics. A minimal model to study the SF process consists of three states with diabats for $\ket{S_0S_1}$, $\ket{T_1T_1}$, and a charge-transfer anion-cation state $\ket{AC}$ which can facilitate the transfer. For a more complete picture, one can include complimentary $\ket{S_1 S_0}$ and $\ket{CA}$ diabats, resulting in a five-state model. To simulate the SF process using our quantum algorithm, the initial state $\ket{\psi(0)} = \ket{S_0 S_1} \bigotimes_{r=0}^{M-1} \ket{\chi_0}$ is prepared as described in \cref{sec:state_prep}. Then, time evolution is performed with the algorithm in \cref{sec:algorithm} yielding $\ket{\psi(t)} = e^{i H t} \ket{\psi(0)}$, from which state populations can be extracted as in Eq.(\ref{eq:population}). By monitoring the population shift from $\ket{S_0 S_1}$ to $\ket{T_1 T_1}$ over time, a quantitative estimate of the SF rate is obtained. Further, state populations could offer valuable insights into the SF mechanism, e.g., the participation of $\ket{AC}$/$\ket{CA}$,~\cite{lukman2016tuning} that practitioners could use to devise improved rational design principles. \\

%Benchmarking
For screening purposes, quantum simulation could be used to study candidate SF chromophores in various ways. For instance, either to answer the decision problem ``Does ultrafast SF occur?" or the more quantitative question, ``What exactly is the transfer rate between the corresponding states?". In general, answering the first question requires fewer resources since one can get away with a lower accuracy in the simulation as we only care if there has been a non-negligible population transfer to the $\ket{T_0 T_1}$ state. For the second question, in contrast, we need a more accurate simulation as we care about the exact population. To illustrate the feasibility of our quantum algorithm within a discovery pipeline, we estimate the implementation cost for a realistic intramolecular SF chromophore. In particular we estimate, based on empirical Trotter error extrapolation, the cost of running our quantum algorithm for $t=100$ femtoseconds on the $N=5$ state, $M=19$ mode model of (NO)$_4$-Anth from Ref.~\cite{Pradhan_Zeng2022} for both a decision type problem at $10\%$ error tolerance and a quantitative problem at $1\%$ error tolerance. {We find they both require  $146$ qubits as well as $\sim 5.47 \times 10^6$ and $\sim 1.73 \times 10^7$ Toffoli gates respectively as shown in \cref{tab:resources}.} This model has been previously studied via non-adiabatic dynamics, which predicted it to have the fastest singlet fission rate~\cite{Pradhan_Zeng2022}. It is, hence, an excellent example of a system that may be probed throughout a molecular design workflow for SF materials. A thorough description of the resource estimation for the reported systems is provided in a Jupyter notebook available at Ref.~\cite{VibeCounting}.

\subsubsection{Triplet separation} 
%Background
Following the generation of a triplet pair $^1(T_1 T_1)$, the triplets might recombine or undergo other dynamical processes that lower the exciton yield. Furthermore, immobile $^1(T_1 T_1)$ triplet pairs are difficult to harvest into charge carriers. A crucial step in retaining the efficiency gain offered by SF is triplet separation,
\begin{align}
\ce{
$^1(T_1 T_1)$ <=> $T_1$ + $T_1$.
}
\end{align}
Due to the general difficulty in its assessment, the rate of triplet separation following SF is a crucial performance metric often left unexplored by current computational screening approaches. %Ref. \cite{pradhan2023triplet} presents the sole study of the quantum dynamics of triplet separation, where the occurrence of triplet separation is assessed for a dimer (NO)$_4$-Anth system. 
A complicating aspect is that the separation of triplets is highly sensitive to the spatial arrangement of chromophores~\cite{johnson2013role, hudson2022next}. Intramolecular SF (iSF) chromophores have been a subject of dedicated interest, in which chromophores are covalently embedded in one molecular frame, allowing for the SF efficiency to be modulated systematically by tuning the covalent linker. In analogy, the triplet separation rate can be tuned by covalent linkage of chromophores. Hence, both SF and triplet separation can be made intramolecular. The concept of intramolecular triplet separation is as novel as iSF was a decade ago and has yet to be systematically explored. Herein, we describe how the general workflow of \cref{fig:workflow} could be used to optimize over covalent linkers for intramolecular triplet separation, graphically presented in \cref{fig:funnels} (B).  After generating a set of candidate covalent frameworks with connected iSF chromophores, one can construct a corresponding vibronic Hamiltonian and monitor the state population transfer from an initial monomeric triplet pair, $\ket{MTP}$, to a dimeric triplet pair \ket{DTP}. The MTP state describes one chromophore in its triplet-pair state $^1(T_1 T_1)$ and the other in its $S_0$ ground state. The DTP state describes the triplet-separated state, where each iSF chromophore now holds an individual triplet state across the covalent bridge. The rate of this state transfer could then be optimized by modifying the chromophore linking. \\

Since intramolecular triplet separation has yet to be systematically explored, the construction of an appropriate diabatic Hamiltonian has yet to be pursued. Ref. \cite{pradhan2023triplet} performs the non-adiabatic dynamics on a vibronic Hamiltonian to describe the \textit{inter}molecular triplet separation for a (NO)$_4$-Anth dimer system with $N=6$ states and $M=21$ vibrational modes. We use a modified version of this vibronic Hamiltonian as a proxy to assess the resource requirements for performing future intramolecular triplet separation simulations using our quantum algorithm. In the original setup, an additional nuclear degree of freedom $R$ is included to describe the distance between monomers, and vibronic interactions are turned on/off using this $R$ dependency. Within the context of intramolecular triplet separation, $R$ would be fixed by a covalent bridge. We hence drop this degree of freedom and replace any $R$-dependent potentials with their values in the $R=3.2$ \AA{} interaction region \cite{pradhan2023triplet} to emulate the intramolecular triplet separation. In \cref{tab:resources}, we report quantum resource estimations for the modified model of (NO)$_4$-Anth dimer triplet separation.

\subsubsection{Charge transfer} \label{sec:charge_transfer}
%Background: 
Following the migration of a separated triplet exciton to the chromophore-acceptor interface, a charge transfer must occur between the excited chromophore, acting as the donor ($D$), and an electron acceptor ($A$), 
\begin{align}
\ce{
$D$ + $A$ <=> $D^{+}$ + $A^{-}$
}
\end{align}
generating a hole-electron pair that gathers at electrodes, transitioning to free charge carriers. To achieve an overall high internal quantum efficiency, triplet excitons residing on the SF chromophores must inject electrons efficiently into the acceptors. Unfortunately, much like the pool of SF chromophores themselves, there is a very limited number of known viable acceptors for SF-based chromophores. Secondly, the most widely studied class of acceptors are fullerenes, which possess poor chemical stability in the presence of acene-based SF chromophores, significantly limiting their applicability~\cite{XiaSanders2017}. There is an urgent need to discover new viable chromophore/acceptor pair systems, which could be accelerated via computational search. Electron transfer processes can occur on ultrafast timescales, involving multiple excited-state pathways along various vibrational degrees of freedom. These systems, hence, often exhibit challenging cases of non-adiabatic dynamics \cite{akimov2014nonadiabatic, xie2015full, di2021multi}, and represent an entire class of industrially relevant simulation instances where our quantum algorithm could provide an advantage over classical methods. \\ 

%Simulation in Pipeline: 
Herein, we describe how our quantum-enabled workflow could be leveraged to design promising chromophore-acceptor interfaces. Similarly to SF and triplet separation, charge transfer is sensitive to the donor-acceptor configuration~\cite{lin2014theoretical}. Of course, chemical reactivity also depends on the spatial arrangement of the two species. One feasible approach to tuning both these parameters is through covalent chromophore-acceptor linking. In principle, given an SF chromophore, both the choice of acceptor molecule and the chromophore-acceptor linking strategy could be optimized. Following the choice of acceptor and covalent chromophore/acceptor bridge, a vibronic Hamiltonian modelling the electron transfer process must be constructed. A minimal model of charge transfer at the chromophore/acceptor interface consists of the photogenerated chromophore excitonic state $\ket{XT}$, and the charge-separated state $\ket{CS}$ ~\cite{di2021multi}. However, a more complete picture would include potentially many intermediate and charge-separated states $\{ \ket{CS_i}  \}_i$, and in the context of donor-bridge-acceptor systems, various intermediate bridge charge-transfer states $\{ CT_j\}_j $ ~\cite{mandal2024ultrafast}. To probe the charge transfer dynamics, an initial state with electronic component in the $\ket{XT}$ state would be prepared and propagated under $e^{i H t}$. The charge transfer process would be monitored via sampling diabatic populations $\ket{CS_i}$ at various time intervals, from which a quantitative charge transfer rate may be fitted. Observing the dynamics through monitoring electronic state populations would potentially offer valuable new information about charge transfer pathways at the chromophore-acceptor interface. \\ 

%Benchmarking
To assess the quantum resource requirements for the simulation of a realistic SF-based chromophore/acceptor electron transfer process, we utilize the anthracene/C$_{60}$ (Anth/C$_{60}$) model described in Ref. \cite{xie2015full}. Fullerene C$_{60}$ is the prototypical acceptor of SF-based solar cells \cite{congreve2013external}, and various anthracene-based SF chromophores have been proposed \cite{pun2018tips, pradhan2020diboron}, making Anth/C$_{60}$ a suitable representative for systems which may be assessed during the quantum-enabled charge transfer optimization step of \cref{fig:funnels} (C). We report the quantum resource estimation in \cref{tab:resources}. To compare our resource estimates with classical methods, Ref.~\cite{xie2015full} required a highly optimized multi-layer MCTDH tree to perform the vibronic dynamics including $N=4$ states and all $M=246$ modes; furthermore, the largest standard MCTDH calculation was limited to $M=11$ selected modes which took 155 CPU hours~\cite{xie2015full}. Furthermore, both the full and reduced models utilized a rather simplified form of vibronic Hamiltonian, where constant interstate couplings $W_{i\neq j} = \lambda^{(i,j)}$ and linear intrastate couplings  $W_{ii} = \lambda^{(i,i)} + \sum_{r} a_r^{(i,i)} Q_r$ are used.  
%interstate ($i \neq j$) couplings $W_{ij}$  were approximated by their zeroeth order contribution, i.e., $W_{i\neq j} = \lambda^{(i,j)}$, and intrastate couplings followed an LVC prescription, $W_{ii} = \lambda^{(i,i)} + \sum_{r} a_r^{(i,i)} Q_r$. 
More accurate parameterizations of the Anth/C$_{60}$ vibronic Hamiltonian, such as a full QVC model or beyond, quickly become prohibitive to address via state-of-the-art classical methods, whereas our quantum algorithm's runtime would still have favorable scaling. Furthermore, modeling the interface charge transfer dynamics could be made even more realistic by including more chromophore or acceptor molecules. The charge transfer dynamics at the chromophore/acceptor interface hence presents a promising future direction for utilizing a quantum computer to address impactful non-adiabatic dynamics simulations. 

\section{Discussion} \label{sec:conclusion}

In this work, we have presented a highly optimized quantum algorithm to implement time evolution under a general vibronic Hamiltonian. Our algorithm is the first in its ability to treat an arbitrary number of electronic states and an arbitrary expansion order of the diabatic potential. Moreover, we develop various algorithmic innovations to reduce the cost of the algorithm, further making it more desirable for implementation. Next, we demonstrate that, unlike many quantum algorithms, our method is not bottlenecked by the input-output problem. Specifically, we show that not only is the cost of preparing relevant initial states negligible but also the key observable in vibronic simulation, electronic state populations, can be extracted by simply measuring a few qubits in the computational basis after time evolution. It's worth noting that we have not addressed the construction of the vibronic Hamiltonian itself (i.e., calculating the coupling parameters that define the model). This often presents its own challenges as it requires electronic structure calculations and diabatization of the potential surfaces. There is, however, progress being made in large-scale excited-state electronic structure~\cite{snyder2016gpu, hollas2018nonadiabatic, mai2020molecular}, automatic diabatization protocols~\cite{zhang2020novel, Neville_Schuurman2024}, and machine learning approaches~\cite{chen2018deep, Chen2019, Kawai_2020, Westermayr2021, ghalami2024nonadiabatic},  which would assist in alleviating this bottleneck. Quantum computers could also be utilized in the electronic structure calculations. \\

Next, we discussed a potential application of our algorithm for industrially relevant problems by outlining a high-level pipeline for the design of singlet fission (SF) solar cells. We showed how our algorithm can be used to probe and optimize for the rates of vibronically driven energy and charge transfer processes important to the overall photovoltaic efficiency. Crucially, the pipeline covers the two important steps following SF: triplet separation and charge transfer. To date, most research in this field has been dedicated to the fission process itself, inadequately leaving the kinetics of the following processes unclear. This pipeline presents an example of where non-adiabatic quantum dynamics could lead to a paradigm shift in the field, fully enabled by the presented quantum algorithm. We note that an economically feasible integration of SF chromophores in organic photovoltaics is also limited by other factors, such as practical synthesis routes, stability under incident light, and resilience against chemical and thermal degradation~\cite{Baldacchino2022}. While our algorithm for dynamics cannot directly assess such additional properties, it can accelerate the discovery of promising SF chromophores and extend the regrettably small pool of available candidates, which is one of the most significant barriers in the development of SF-based solar cells~\cite{Casanova2018, padula2019singlet, Pradhan_Zeng2022, liu2022finding, zeng2014seeking}. 

\section{Acknowledgements}
The authors thank Ilya G. Ryabinkin and Michael Schuurman for stimulating discussions. We also thank Zhenggang Lan and Jiawei Peng for providing the vibronic Hamiltonian parameters for the anthracene/C$_{60}$ electron transfer model from Ref. \cite{xie2015full} used in \cref{sec:design_pipeline}.
A.A.-G. thanks Anders G. Frøseth for his generous support, and acknowledges the generous support of Natural Resources Canada and the Canada 150 Research Chairs program.
T.Z. thanks the Natural Sciences and Engineering Research Council (NSERC) of Canada (RGPIN-2024-06286) for financial support and the Digital Research Alliance of Canada for computational resources.

\bibliography{main}
% %\printbibliography
\appendix

\section{Complexity Analysis}\label{ap:proof}
\marktext{
Here we provide a proof for the total cost of time evolution presented using the quantum algorithm presented in \cref{sec:algorithm}.  
\begin{proof}[Proof of \cref{thm:complexity}]
    Cost of higher order Trotter product formulas in terms of the cost of first order Trotter is well established \cite{papageorgiou2012efficiency, childs2021theory}. Therefore, we focus on the cost of a first order Trotter step, $\Omega$, here. We seek to prove
    \begin{equation}
        \Omega \in O(M^d(N\cdot d\cdot\log^2(K) + N^2 )).
    \end{equation}
    Further, we wish to show that the presented fragmentation scheme uses $O(d^2\cdot \log(K)+\log(N))$ ancilla qubits. For the following, we assume the polynomial coefficients are given as floating-point numbers of constant lengths. Note that the cost of a single first-order Trotter step reduces to that of implementing the exponentials of all fragments
    \begin{equation}
        \Omega = \# \text{fragments}\,\cdot (\text{Cost(diagonalization)} + \text{Cost(diagonal operator))}.
    \end{equation}
    The cost of implementing $e^{i T}$ will be $2M$ times the cost of $\log(K)$-qubit QFTs, $M$ squaring of $\log(K)$-qubit registers, and $M$ multiplications of $2\log(K)$-qubit registers with floating-point numbers of constant lengths for a total non-Clifford cost of $O\left(M\log(K)^2\right)$. Hence, for the rest of the proof, we'll focus on the cost of implementing the potential.\\
    The fragmentation scheme results in $N$ fragments where the non-Clifford cost of diagonalizing each fragment is zero. Implementing a diagonalized fragment requires exponentiating $O(M^d)$ terms of $d^{th}$ degree, giving
    \begin{equation}
        \Omega \in O(N \cdot M^d\cdot \text{Cost(exponentiating a $d^{th}$ degree term)}).
    \end{equation}
    Hence it suffices to show Cost(exponentiating a $d^{th}$ degree term) is $O(d\log^2(K)+N)$ Toffolis and uses at most $O(d^2\cdot \log(K)+\log(N))$ ancilla qubits. To exponentiate a term of $d^{th}$ degree with different coefficients for different electronic states, we load the electronic state-dependent coefficients in an ancilla register of fixed length using QROM. This uses $O(N)$ Toffolis and $O(\log(N))$ ancillas for the unitary iterator. We then need to compute the $d^{th}$ degree variable associated to the term using quantum arithmetic. Using the caching algorithm, we will always have a corresponding $d-1$ degree term cached in a register of size $(d-1)\log^2(K)$, to be multiplied with a $1^{\rm{st}}$ degree term to get our $d$ degree term using $O(d\log^2(K))$ Toffolis. If the caching technique was not employed, and we had to compute every term from scratch, we would need $O(d^2\log^2(K))$ Toffolis. Next, adding the product of the $d$ degree variable and the coefficient register to a phase gradient register of the same fixed length as the coefficient also requires $O(d\log^2(K))$ Toffolis. Lastly, keeping a cached variable of each degree up to $d$ requires $O(d^2\log(K))$ ancilla qubits (note that even without caching we'd still need the same number of ancillas for computing a $d$ degree term from scratch). Hence, in total we require $O(d\log^2(K)+N)$ Toffolis and $O(d^2\cdot \log(K)+\log(N))$ ancilla qubits to exponentiate a $d^{th}$ degree term as desired.
\end{proof}
}

{
\section{Comparison to classical methods} \label{app:classical_hardness}
Herein we provide a brief discussion of classical methods for performing fully quantum non-adiabatic dynamics, along with the factors limiting their effectiveness in the context of a computational materials discovery pipeline. Classical methods for quantum dynamics under the vibronic Hamiltonian Eq.(\ref{eq:KDC_Hamiltonian}) typically utilize the insertion of an ansatz for the vibronic wavefunction into a time-dependent variational principle (TDVP). For simplicity, we can initially consider the adiabatic case of $M$ vibrational modes and a single electronic state. In this case, the fully expressive, numerically exact ansatz for the time-dependent wavefunction may be written
\begin{align} \label{eq:classical_exact}
\psi(Q_1, \hdots, Q_M, t) = \sum_{p_1, \hdots, p_M=1}^{K} C_{p_1, \hdots p_M} (t) \prod_{r=1}^M \chi_r^{(p_r)} (Q_r) ,
\end{align}
where each mode is given a basis of $K$ primitive time-independent functions or grid-points in analogy to the quantum algorithm. Insertion of Eq.(\ref{eq:classical_exact}) into a TDVP provides equations of motion on time-dependent tensor $C_{p_1, \hdots p_M} (t)$,
\begin{align} \label{eq:classical_exact_eoms}
\frac{d}{dt} C_{p_1, \hdots p_M} = \sum_{q_1 \hdots q_M = 1}^{K} C_{q_1 \hdots q_M} \bra{ \chi_1^{(p_1)} \hdots \chi_M^{(p_M)}} H \ket{ \chi_1^{(q_1)} \hdots \chi_M^{(q_M)}}. 
\end{align}
The memory requirements of storing the wavefunction parameters in Eq.(\ref{eq:classical_exact}) scale as $K^{M}$, and the number of operations required for performing a single step of time integration through Eq.(\ref{eq:classical_exact_eoms}) scales as $M K^{M+1}$ if the Hamiltonian is written in a sum-of-products form \cite{beck2000multiconfiguration}. To circumvent the exponential scaling in the size of the primitive basis, the MCTDH method introduces $K^{'} < K$ time-dependent basis functions referred to as single particle functions (SPFs) $\phi_r^{(p_r)}(Q_r, t)$, and the MCTDH ansatz can be written
\begin{align} \label{eq:classical_mctdh}
\psi_{\rm{MCTDH}}(Q_1 \hdots Q_M, t) = \sum_{p_1, \hdots p_M = 1}^{K^{'}} A_{p_1, \hdots p_M} (t) \prod_{r = 1}^{M} \phi_r^{(p_r)}(Q_r, t).
\end{align}
By expanding the wavefunction in $K^{'}$ SPFs rather than  $K$ primitive basis function, storing and updating the MCTDH wavefunction requires memory and time scaling dominated by $(K^{'})^M$ and $M^2(K^{'})^{M+1}$ respectively. Utilizing a basis set of SPFs adaptively changing over time offers the possibility of using fewer basis functions to describe the dynamics, however it is inherently an approximation unless $K^{'}=K$. The time integration can become less accurate at further times, particularly when the dynamics become chaotic \cite{Manthe1992}. Lowering the base in the time and memory scalings allows MCTDH to approximately handle larger systems than with the standard numerically exact procedure, however it is still fundamentally limited by the growth of entanglement as the system propagates. To extend Eq.(\ref{eq:classical_exact}) or Eq.(\ref{eq:classical_mctdh}) to non-adiabatic systems involving $N$ electronic states, each diabatic state is given a vibrational counterpart taking the form of Eq.(\ref{eq:classical_exact}) or Eq.(\ref{eq:classical_mctdh}),
\begin{align}
\ket{\Psi(t)} =  \sum_{i=1}^N |\psi^{(i)}(t) \rangle \otimes \ket{i}.
\end{align}
The exact memory and time scalings of MCTDH in the non-adiabatic case are more complicated, since they depend on if the so-called single-set or multi-set formalisms are used \cite{beck2000multiconfiguration}. However, they are generally characterized by at least linear $N$ dependence, and retain the $(K^{'})^{M}$ and $M^2 (K^{'})^{M+1}$ dependencies respectively. Mode combinations can be used to mitigate the leading contributions to the memory and time requirements, by replacing e.g., $M^2 (K^{'})^{M+1}$ with $(M^{'})^2 (\tilde{K})^{M^{'}+1}$, where  $M'$ is the number of combined modes and $\tilde{K}$ is the number of multimodal functions used per combined mode. As the number $M/M^{'}$ of physical modes per combined mode increases, the dominant computational cost shifts to description of the multimodal functions due to exponential scaling in $\frac{M}{M^{'}}$. In the multi-layer MCTDH (ML-MCTDH), combined modes are further combined hierarchically forming a ``multi-layer tree" \cite{wang2003multilayer}, which can be viewed as representing a hierarchical Tucker tensor decomposition of $C_{p_1 \hdots p_M}(t)$ in Eq.(\ref{eq:classical_exact}) \cite{Larsson2024}. This high degree of flexibility in the ansatz allows for careful balancing between the sources of exponential cost, however finding an appropriate hierarchical representation and demonstrating its convergence can be extremely challenging tasks in practice \cite{Welsch2012, Larsson2024}. This feature makes ML-MCTDH unfavorable within the context of a computational material discovery pipeline, since a potentially complicated multilayer tree must be made in a system-specific fashion for each candidate. Furthermore, vibronic Hamiltonians with higher order terms such as  bilinear $Q_r Q_s$ and trilinear $Q_r Q_s Q_t$ interactions are expected to more rapidly produce correlations between the vibrational modes, making the identification of an appropriate tensor decomposition considerably more challenging, even for short time dynamics. Similar difficulties would be expected for other tensor network methods besides ML-MCTDH, such as the time-dependent density matrix renormalization group \cite{tddmrg1}. We hence expect quantum advantage in vibronic dynamics to be found in problem instances with rapid entanglement formation across many degrees of freedom, where higher-order interactions must be included to accurately predict experimental outcomes. Sophisticated classical tensor network approaches face a substantial increase in resources to accurately represent and propagate highly correlated vibronic wavefunctions. On the other hand, the quantum algorithm presented in this work benefits from efficient scaling in $N$, $K$, and $M$ with no dependence on entanglement structure.}

\end{document}